\documentstyle[11pt,twoside,paspconf]{article}

\begin{document}

\title{Accretion history of super-massive black holes}
\author{Priyamvada Natarajan}
\affil{Canadian Institute for Theoretical Astrophysics, 60 St. George
Street, Toronto M5S 3H8, Canada}

\begin{abstract}
We show that the luminosity function of the actively star-forming
Lyman break galaxies and the B-band quasar luminosity function at $z =
3$ can be fit reasonably well with the mass function of collapsed
galaxy scale dark matter haloes predicted by viable variants of
hierarchical cold dark matter dominated cosmological models for
lifetimes $t_Q$ of the optically bright phase of QSOs in the range
$10^{6}$ to $10^{8}$ yr. There is a strong correlation between $t_Q$
and the required degree of non-linearity in the relation between black
hole and host halo mass. Such a non-linear relation is motivated by
suggesting that the mass of supermassive black holes may be limited by
the back-reaction of the emitted energy on the accretion flow in a
self-gravitating disc. This would imply a relation of black hole to
halo mass of the form $M_{\rm bh} \propto v_{\rm halo}^5 \propto
M_{\rm halo}^{5/3}$ and a typical duration of the optically bright QSO
phase of the order of the Salpeter time, $\sim\,10^{7}$ yr. The high
integrated local mass density of black holes inferred from recent
kinematic determinations of black hole masses in nearby galaxies seem
to indicate that the overall efficiency of supermassive black holes
for producing blue light is lower than was previously assumed.  We
discuss three possible accretion modes with low optical emission
efficiency: (i) accretion well above the Eddington rate, (ii)
accretion obscured by dust, and (iii) accretion below the critical
rate leading to an advection dominated accretion flow lasting for a
Hubble time.  We further argue that accretion with low optical
efficiency might be closely related to the origin of the hard X-ray
background.
\end{abstract}


\section{\bf Motivation}

Bright QSO activity in the optical peaks at around redshift $z$=2.5
(Schmidt, Schneider \& Gunn 1994; Warren, Hewett \& Osmer 1994;
Shaver et al.~1996; McMahon, Irwin \& Hazard 1997). QSOs are believed
to be powered by the accretion onto super-massive black holes at the
centre of galaxies (e.g. Rees 1984 and references therein) and a
number of authors have linked the changes in QSO activity to changes
in the fuel supply at the centre of the host galaxies (Cavaliere \&
Szalay 1986; Wandel 1991; Small \& Blandford 1992).  Haehnelt
\& Rees (1993; HR93 henceforth; see also Efstathiou \& Rees 1988)
recognized that the peak of QSO activity coincides with the time when
the first deep potential wells assemble in plausible variants of
hierarchical cosmogonies.  The past few years have seen dramatic
observational improvements in the detection of galaxies out to $z >
$3, transforming our knowledge of galaxy and star formation in the
high redshift universe. There are also far more extensive data on the
demography of supermassive black holes in nearby galaxies, and on low
level activity of AGN both in the optical and the X-ray bands. We
discuss here some implications for the formation and evolution of
active galactic nuclei, attempting to tie these lines of evidence
from high and low redshifts together in a consistent model.

\section{\bf Relating the luminosity function of QSOs and
star-forming galaxies to the mass function of dark matter (DM) haloes}

Steidel and collaborators (Steidel et al. 1992; Steidel et al. 1995;
Steidel et al. 1996; Giavalisco et al. 1997) have developed a
technique for picking out galaxies in the redshift range
$2.5\,<\,z\,<\,3.5$. The detected star-forming population at $z = 3$
bears a close resemblance to local star-burst galaxies. The abundance
of these objects is found to be roughly half of that of $L\,>\,{L^*}$
present day galaxies. There are however {\bf no secure} direct
dynamical mass estimates for these Lyman-break galaxies. And little is
known at present about the the relation of the masses of these objects
to the rate of detected star formation. Strong clustering detected in
these Lyman-break systems at $z\,\sim\,3$ leads to an interpretation
of these objects as the potential progenitors of massive galaxies at
the present epoch.

Several groups have been engaged in the quest for high-redshift
quasars. Schmidt, Schneider and Gunn (1994), detected 90 quasars by
their Lyman-$\alpha$ emission from the Palomar Transit Grism Survey in
the redshift range $2.75 - 4.75$. They find that the space density of
$M_B\,<\,-26$ quasars decreases by a factor of 2.7 per unit redshift
beyond $z = 2.7$. Based on their analysis they conclude that the peak
of the co-moving space density distribution of quasars with
$M_B\,<\,-26$ lies between redshifts $1.7 - 2.7$.  The observed
decline in the QSO number density could be either due to the real
paucity of high-z QSOs or due to obscuration by dust introducing a
systematic bias at high redshifts. In a recent paper, Shaver et. al
(1996) report the results of the Parkes survey of a large sample of
flat spectrum sources in the radio-waveband (that is unaffected by
dust) covering roughly 40\% of the sky.  They claim that the
space-density of radio-loud quasars does indeed decline with redshift
at $z\,>\,3$, and argue that the same conclusion probably applies to
all quasars. 

In what follows, we explore the link between star-forming galaxies and
QSOs at high redshift, assuming that both populations trace the mass
function of DM haloes. Using the Press-Schechter formalism we obtain
estimates of the space density of DM haloes. The space density of
these star-forming galaxies corresponds to those of haloes with masses
of $~10^{12.5}\,M_{\odot}$ and virial velocities of $~300\,$km
s$^{-1}$ (see also Baugh et al.~1997).  Further evidence for masses of
this order comes from the strong clustering of these galaxies (Steidel
et al.~1997, Bagla 1997, Jing \& Suto 1997, Frenk et al.~1997, Peacock
1997).  A reasonable fit to the luminosity function of these objects
can be obtained assuming a linear relation between star formation rate
and halo mass, i.e. a constant mass-to-light ratio. A weakly non-linear
relation is also consistent with the data and would indeed be required
to match the shallow slope of the luminosity function at the high
luminosity end reported recently by Bershady et al. (1997). Comparable
fits are obtained for the other CDM variants. Note, however, that the
observed $H_{\beta}$ widths of these galaxies, $\sigma \sim 80\,$km
s$^{-1}$ (Pettini et al.~1997), might be in conflict with the virial
velocities of $300\,$ km s$^{-1}$ quoted above.

The space density of optically selected QSOs at $z\,=\,$3 with $M_{\rm
B} <-23$ is smaller than that of the detected star-forming galaxies by
a factor of a few hundred. As demonstrated by HR93 the
well-synchronized evolution of optically selected QSOs can be linked
to the hierarchical growth of DM haloes on a similar timescale if the
duration $t_{\rm Q}$ of the optically bright phase is considerably
shorter than the Hubble time. For small $t_{\rm
Q}$ this comes more and more in line with the predicted space density
of DM haloes and that of star-forming galaxies at high
redshift. However, hardly anything is known about the masses of the
host objects of optically selected QSOs and this still leaves
considerable freedom in the exact choice of $t_{\rm Q}$. Following the
approach of HR93 we estimate the formation rate of active black holes
by taking the positive term of the time derivative of the halo mass
function and a simple parameterization for the black hole formation
efficiency.  It is further assumed that active black holes radiate
with a light curve of the form, $L_{\rm B}(t) = f_{\rm B}\,f_{\rm Edd}
\, L_{\rm Edd}\exp{(-t/t_{\rm Q})}$, where $f_{\rm Edd}$ is the ratio
of bolometric to Eddington luminosity and $f_{B}$ is the fraction of
the bolometric luminosity radiated in the B-band.

\begin{figure}
\vspace{12truecm}
\caption{Model A: The B-band QSO luminosity function at $z = 3$ for 2
cosmological models computed using the time derivative of the space
density of DM halos. A bolometric correction factor of 6.0. In the 
lower panel: a non-linear relation with $\alpha = 5/3$ between the 
BH mass and the halo mass with a QSO lifetime of $1 \times 10^7\,$ yr 
was assumed and in the upper panel a linear relation between
the accreting BH mass and the DM halo mass with a shorter lifetime 
for the QSO was assumed $1 \times 10^6$ yr [the same parameters as the
best-fit model explored in Haiman \& Loeb 1997]. The over-plotted 
data points are from Boyle et al. (1988).}
\end{figure}

We obtain reasonable fits for a wide range of lifetimes and for all
the CDM variants if we allow ourselves some freedom in the relation
between {\bf halo mass} and {\bf black hole mass}.  There are,
however, systematic trends: with increasing lifetime the black hole
mass has to become a progressively more nonlinear function of the halo
mass and the black hole formation efficiency has to decrease in order
to match the luminosity function of QSOs. This is due to the fact that
QSOs are identified with rarer and more massive haloes with increasing
lifetime, and these fall on successively steeper portions of the halo
mass function. In Fig. 1 we plot two specific choices of parameters
which we denote as model A and B hereafter (see Haehnelt, Natarajan \&
Rees (1998) for more details). In the lower panels a QSO lifetime close
to the Salpeter timescale $t_{\rm Salp}\,= \epsilon \sigma_{\rm T}
c/4\pi G m_{\rm p} = \,4.5\, \epsilon_{0.1}\, 10^{7}$yr and a scaling
of black hole mass with halo virial velocity as $M_{\rm bh}\,\propto\,
v_{\rm halo}^{5}\propto M_{\rm halo}^{5/3} (1+z)^{5/2}$ was assumed
($\epsilon$ is the total efficiency for transforming accreted rest
mass energy into radiation). A physical motivation for this particular
dependence is discussed later. The upper panel shows the case of a
linear relation between the halo and black hole mass advocated by
Haiman and Loeb (1997,HL97) which requires a QSO lifetime of less than
$10^{6}$yr -- much shorter than the Salpeter time for usually assumed
values of $\epsilon$. In principle $t_{Q}$ could also depend on mass
or other parameters.

\section{\bf Local demography of black holes}

The last few years have seen tremendous progress in establishing the
existence of supermassive black holes, there are now a number of
excellent cases (including that of our own Galaxy) where observations
strongly imply the presence of a relativistic potential well (Watson
\& Wallin 1994; Miyoshi et al.~1995; Genzel et al.~1997). Magorrian
et al.  (1997; Mag97 henceforth) published a sample of about thirty
estimates for the masses of the putative black holes in the bulges of
nearby galaxies.  Mag97 confirm previous claims of a strong
correlation between bulge and black hole mass (Kormendy \& Richstone
1995).  A linear relation of the form, $M_{\rm bh}= 0.006\,M_{\rm
bulge}$, was obtained by Mag97 as a best fit. However, considering the
large scatter a mildly non-linear relation would probably also be
consistent with the data. We would further like to note here that a
linear relation between black hole to bulge mass does not necessarily
imply a linear relation between black hole and halo mass and as we
will argue later a non-linear relation might be more plausible.
Fugukita, Hogan \& Peebles (1997) estimate the total mass density in
stellar bulges as $ 0.001h^{-1} \le \Omega_{\rm bulges} \le
0.003h^{-1}$ and together with the above ratio of black hole to bulge
mass we get,
$$
{\rho_{\rm bh}}=3.3h\times 10^{6}\,({M_{\rm bh}/M_{\rm bulge}}/{0.006})
\,({\Omega_{\rm bulge}}/{0.002h^{-1}})\,M_{\odot}\,Mpc^{-3}.  
$$
Considering the complicated selection biases of the Mag97 sample, the
small sample size and possible systematic errors in the black hole
mass estimates this number is still rather uncertain. Van der Marel
(1997) e.g.  emphasizes the sensitivity of black hole mass estimates
to the possible anisotropy of the stellar velocity distribution and
argues that the Mag97 mass estimates might be systematically too high.
Nevertheless, as pointed out by Phinney (1997; see also Faber et
al.~1996) $\rho_{\rm bh} ({\rm nearby\ galaxies})$ exceeds the mass
density in black holes needed to explain the blue light of QSOs purely
by accretion onto super-massive black holes,
$$
\rho_{\rm acc} ({\rm QSO}) = 1.4\times 10^{5}\,({f_{\rm B}\,
\epsilon}/{0.01})^{-1}\,M_{\odot}\,Mpc^{-3},
$$
by a factor of about ten unless the value of $f_{\rm B}\, \epsilon$ is
smaller than usually assumed (Soltan 1982, Chokshi \& Turner 1992).
While a few years ago it seemed difficult to discover the total mass
in black holes necessary to explain the blue light emitted by QSOs at
high redshift, black hole detections in nearby galaxies now suggest
that accretion onto supermassive black holes may actually be rather
inefficient in producing blue light.

\section{\bf Constraints on the accretion history 
of supermassive black holes}

There are three options to explain the apparently large value of
$$\rho_{\rm bh} ({\rm nearby\ galaxies})/\rho_{\rm acc} ({\rm QSO})$$
(i)$\rho_{\rm bh} ({\rm nearby\ galaxies})$ is strongly
overestimated, or (ii) $f_{B}\,\epsilon$ during the optically bright
phase is smaller than previously assumed, or (iii) supermassive black
holes do not gain most of their mass during the optically bright
phase. A plausible solution with $f_{B}\,\epsilon$ significantly
smaller than 0.01 is discussed later in this section. We first explore
the third possibility somewhat further. The typical mass of a black hole
at the end of the optically bright phase of duration $t_{\rm Q}$
exceeds that accreted during this phase by a factor $M_{\rm bh}/M_{\rm
acc} = f_{\rm edd}^{-1}\, t_{\rm Salp}/t_{\rm Q}$.  This factor should
be larger than 1 and smaller than $\rho_{\rm bh} ({\rm nearby\
galaxies}) /\rho_{\rm acc} ({\rm QSO})$ and therefore,
$$
1\,le\,f_{\rm Edd}^{-1}\,\epsilon^{-1}\,({t_{\rm Q}}/{4.5\times
10^{8}\,yr}) ^{-1}\,\le\,{25h ({f_{B}\,\epsilon}/{0.01})\,(
{\rho_{\rm bh}}/{3.3h\times 10^{6}\,Mpc^{-3}}}).
$$

The question when supermassive black holes gained most of their mass
is therefore closely related to $t_{\rm Q}$ and $f_{\rm Edd}$.  For
bright quasars, $f_{\rm Edd}$ must be $> 0.1$; otherwise excessively
massive individual black holes would be required.  Furthermore,
$f_{\rm Edd}$ will always be smaller than unity even if the ratio of
the accretion rate to that necessary to sustain the Eddington
luminosity, $\dot m$, greatly exceeds unity. This is because a
``trapping surface'' develops at a radius proportional to $\dot m$,
within which the radiation advects inwards rather than escapes.  In
consequence, the emission efficiency declines inversely with $\dot m$
for $\dot m >1$ (Begelman 1978).

For $0.1 \le f_{\rm Edd} \le 1$  the possible range 
for $t_{\rm Q}$  is, 
$$
{2h^{-1}\times 10^{6}({f_{\rm B}}/{0.1})^{-1}\,({\rho_{\rm bh}/{3.3h\times
 10^{6} \,Mpc^{-3}}})^{-1}\,yr}\,\le\,t_{\rm Q}\,\le 4.5 
\times 10^{8}\,({\epsilon}/{0.1})\,yr.  
$$

If $t_{\rm Q}$ is very short (as in model B) and $f_{B}\,\epsilon$ is
not significantly smaller than 0.01 it seems inevitable that
supermassive black holes have acquired most of their mass {\it before}
the optically bright phase. We would like to point out here that a
value of $\rho_{\rm bh}$ as large or larger than we infer from Mag97
is actually needed for short $t_{\rm Q}$.

For the remainder of this section we assume that $t_{\rm Q}$ is of
order the Salpeter time. The ratio of accreted mass to total mass at
the end of the optically bright phase is then equal to $f_{\rm
Edd}^{-1}$.  If $f_{\rm Edd} \sim 1$ during the optically bright phase
(and if $f_{B}\,\epsilon$ is not significantly smaller than 0.01) then
the corresponding gain in mass by a factor $\rho_{\rm bh} ({\rm
nearby\ galaxies})/\rho_{\rm acc} ({\rm QSO})$ indicated by Mag97 has
to occur {\it after} the optically bright phase.  As the accretion
should not be optically bright the most plausible options are
advection dominated accretion flows (ADAFs) and dust-obscured
accretion (Narayan \& Yi 1995, Fabian et al.~1997 and earlier
references cited therein). ADAFs require low accretion rate with $\dot
m < m_{\rm crit}$ where $\dot m_{\rm crit} = 0.3\alpha_{\rm ADAF}^{2}$
and $\alpha$ is the Shakura-Sunyaev viscosity parameter. There is
therefore a maximum growth factor for the black hole mass density due
to ADAFs $\sim 3.0\alpha_{\rm ADAF}^{2} t_{\rm ADAF}/t_{\rm
Salp}(\epsilon = 1)$ and \hfill \break $\alpha_{\rm ADAF}>
0.3\,[\epsilon\, \rho_{\rm bh} ({\rm nearby\ galaxies})/\rho_{\rm acc}
({\rm QSO})]^{0.5}$ would be required even if the accretion lasts all
the way from $z$=3 to $z$=0.

If, however, $f_{\rm Edd} \sim 0.1$ (and $t_{\rm Q} \sim t_{\rm
Salp}$)then the gain in mass by a factor $\rho_{\rm bh} ({\rm nearby\
galaxies})/\rho_{\rm acc} ({\rm QSO})$ has to occur {\it before} the
optically bright phase as in the case of small $t_{\rm Q}$.  For ADAFs
this would require $\alpha_{\rm ADAF} \sim 1$ and is therefore hardly
plausible. In this case dust-obscured accretion would be the only
viable option.

\section{\bf Possible accretion histories with low optical efficiency}

\begin{figure}
\vspace{12truecm}
\caption{Two accretion histories with low overall optical
emission efficiencies are illustrated here.}
\end{figure}

Two possible accretion histories with a low overall efficiency for
producing blue light are sketched in Fig. 3 - the solid curves describe an
accretion history where most of the mass is accreted during the ADAF
phase while the dashed curves are for an accretion history where the
black hole gains most of its mass during a short-lived early phase
with with $\dot m>1$. The figure shows mass accretion rate in units of
the Eddington accretion rate, the mass relative to the final mass and
the optical and hard X-ray luminosity.  The accretion rate is constant
at the beginning with $\dot m >1$. The mass is therefore linearly
rising and $\dot m$ decreases. The spectral energy distribution for
accretion with $\dot m>1$ is rather uncertain (as indicated by the
three parallel lines for $\dot m > 1$ in the two bottom panels) and
should depend on the absorbing column and the dust content of the
outer parts of the self-gravitating disc and/or the host galaxy.  The
sharp drop of $\dot m$ marks the onset of the back-reaction on the
accretion flow and either the start or the peak of the optical bright
phase (with a rather inefficient production of hard X-rays).  Once the
accretion rate has fallen below the critical rate for an ADAF
(indicated by the dashed lines in the top panel) the spectral energy
distribution will change to one peaked in the hard X-ray waveband.

\section{\bf Faint X-ray sources and the hard X-ray background}

As pointed out by many authors, the X-ray emission of optically
selected QSOs is too soft to explain the hard X-ray background.  Di
Matteo \& Fabian (1996) and Yi \& Boughn (1997) argued that the
emission from ADAFs has a spectral shape similar to the hard X-ray
background. Fabian et al.~(1997) suggested that this might also be
true for dust-obscured accretion. It is therefore tempting to link the
rather large value of $\rho_{\rm bh} ({\rm nearby\ galaxies})
/\rho_{\rm acc} ({\rm QSO})$ inferred from Mag97 to the origin of the
hard X-ray background and the recently detected large space density of
faint X-ray sources (Almaini et al.~1996; Hasinger et al.~1997,
Schmidt et. al 1997, McHardy et al.~1997, Hasinger 1998). The presence
of extremely low-level optical AGN activity in a large fraction of
galaxies reported by Ho, Fillipenko \& Sargent (1997) would also fit
in nicely with such a picture. The efficiency of ADAFs is
$\epsilon_{\rm ADAF} =0.1\, (\alpha/0.3)^{2}\, \dot m/\dot m_{\rm
crit}$ and decreases rapidly for small $\alpha$ and small $\dot m/\dot
m_{\rm crit}$.  If the hard-Xray background was produced by ADAFs onto
supermassive black holes in ordinary galaxies this requires a value of
$\rho_{\rm bh} ({\rm nearby\ galaxies})$ as high as we infer from
Mag97, a large value of $\alpha$ and a value of $\dot m$ below but
still close to $\dot m_{\rm crit}$ lasting for a Hubble time for the
majority of supermassive black holes. At faint flux levels and high
redshifts a possible star-burst contribution to the total spectral
energy distribution will become more and more important in the optical
and probably also the soft X-ray. It is interesting to note here that
the recently detected number counts in the sub-mm wave-band by SCUBA
could be evidence for dust-obscured accreting AGN at high redshifts
(Blain et al. 1998; Almaini et al. 1998).

\section{\bf Conclusions}

The optical QSO luminosity function at $z\sim 3$ can be plausibly
matched with the luminosity function of star forming galaxies at the
same redshift and the mass function of DM haloes predicted by a range
of variants of CDM cosmogonies believed to comply with observational
constraints in the low-redshift universe.  This is possible for
lifetimes of optically bright QSOs anywhere in the range $10^{6}$ to
$10^{8}\,yr$.  There is a correlation between the lifetime and the
required degree of non-linearity in the relation between black hole
and halo mass.  The non-linearity has to increase for increasing
lifetime. Predicted host halo masses, host galaxy luminosities, and
the clustering strength all increase with increasing lifetime and
further observations of these offer our best hope of constraining the
duration of the optically bright phase of QSOs.

The present-day black hole mass density implied by the integrated
luminosities of optically bright QSO may be significantly smaller than
that inferred from recent black hole estimates in nearby galaxies for
generally assumed efficiencies for producing blue light.  We have
discussed three possibilities for how and when this mass could be
accreted in an optically inconspicuous way: (i) in the early stages of
accretion at rates far above the Eddington rate, (ii) by accretion
where optical emission is obscured by dust, or (iii) in the late
stages of accretion at a rate below the critical rate for an advection
dominated accretion flow with an Shakura-Sunyaev parameter of
$\alpha_{\rm ADAF}>0.3$.  

\acknowledgments

The Isaac Newton Institute for Mathematical Sciences and specifically
the program on the Dynamics of Astrophysical Discs are gratefully
acknowledged for providing a lively scientific environment. Thanks
are due to my collaborators Martin Haehnelt and Martin Rees for
permission to present results from our joint work.

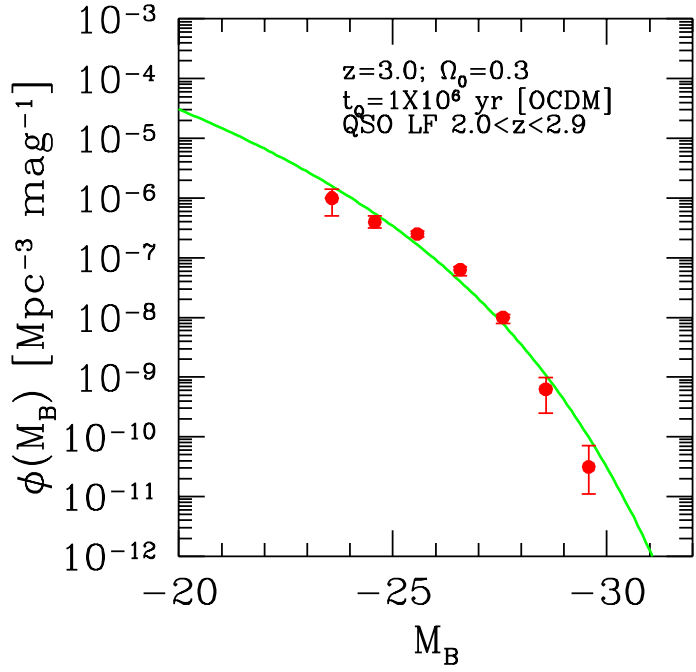
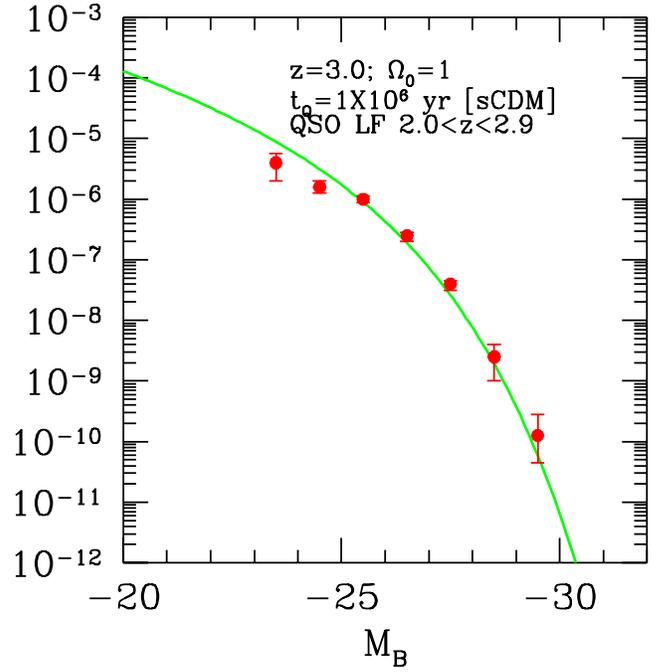
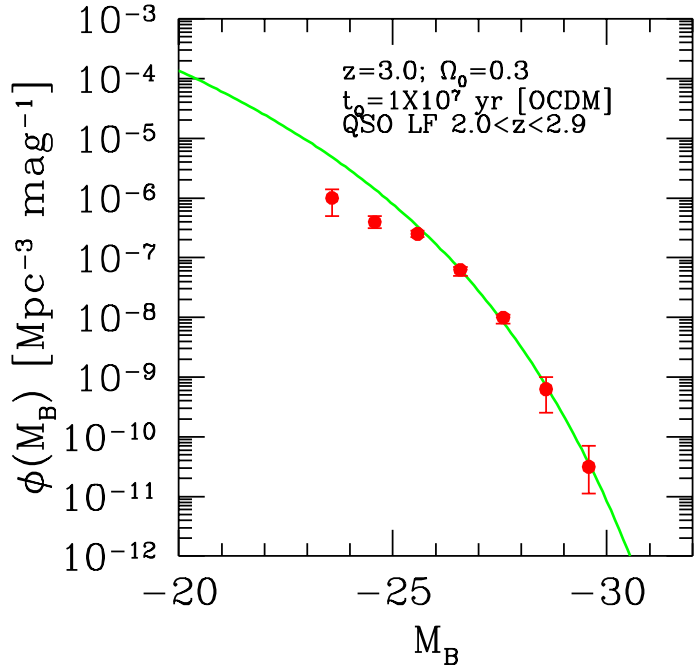
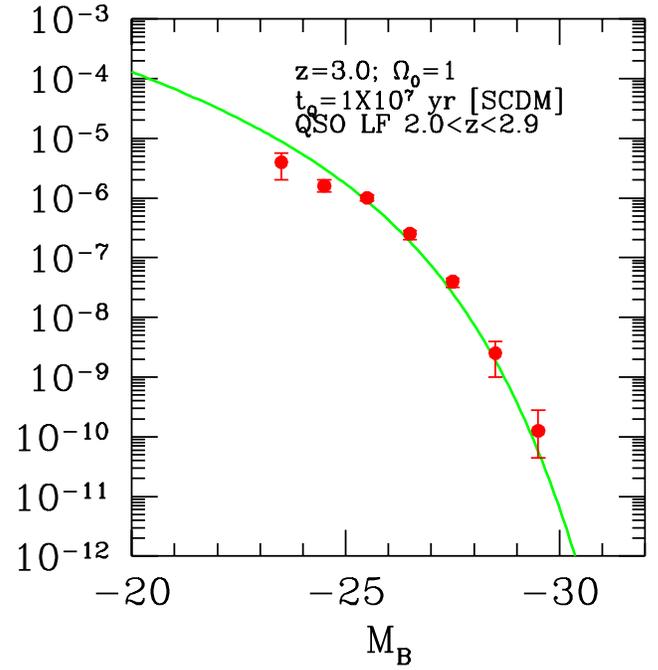
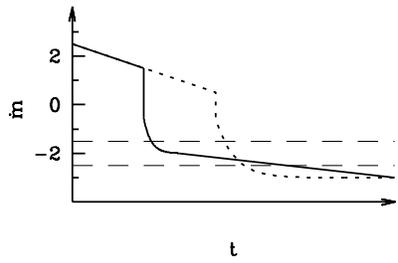
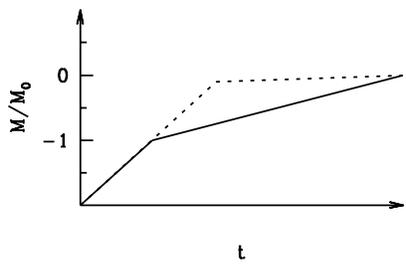
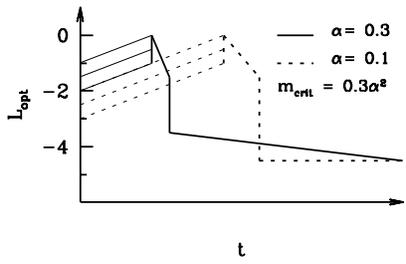
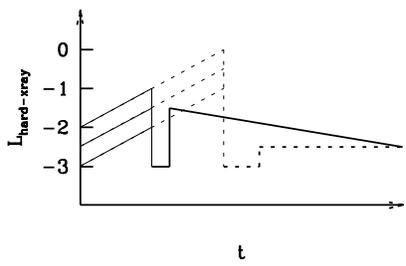
\end{document}